\documentclass[journal]{IEEEtran}
\newcommand{\wv}{{\bf w}}

\newcommand{\xv}{{\bf x}}
\newcommand{\yv}{{\bf y}}

\newcommand{\Am}{{\bf A}}

\newcommand{\Hc}{{\cal H}}
\ifCLASSINFOpdf
\else
\fi

\usepackage{graphicx,picture}
\usepackage{bm}
\usepackage{amsfonts}
\usepackage{amsmath}
\usepackage{amssymb}
\usepackage{geometry}
\usepackage{times}
\usepackage{cite}
\usepackage{subfigure}
\usepackage{latexsym,bm,amsmath,amssymb} 
\usepackage{CJK}
\usepackage{hhline}

\usepackage{multirow} 
\usepackage{amsmath}
\usepackage{xcolor}
\usepackage[linesnumbered,ruled,vlined]{algorithm2e}
\usepackage{epstopdf}
\usepackage{url}
\begin{document}

%
\title{From OTFS to DD-ISAC: Integrating Sensing and Communications in the Delay Doppler Domain}
\author{Weijie Yuan, \IEEEmembership{Member, IEEE,} Lin Zhou, \IEEEmembership{Member, IEEE}, Saeid K. Dehkordi, \IEEEmembership{Student Member, IEEE}, Shuangyang Li, \IEEEmembership{Member, IEEE}, Pingzhi Fan, \IEEEmembership{Fellow, IEEE}, Giuseppe Caire, \IEEEmembership{Fellow, IEEE}, and H. Vincent Poor, \IEEEmembership{Life Fellow, IEEE}

\thanks{This work was supported in part by the U.S National Science Foundation under Grant CNS-2128448.

W. Yuan is with the Department of Electrical and Electronic Engineering, Southern University of Science and Technology, Shenzhen 518055, China 
(e-mail:yuanwj@sustech.edu.cn).

L. Zhou is with the School of Cyber Science and Technology and the Beijing Laboratory for General Aviation Technology, Beihang University, Beijing 100191, China (e-mail: lzhou@buaa.edu.cn).

S. K. Dehkordi, S. Li, and G. Caire are with the Department of Electrical Engineering and Computer Science, Technical University of Berlin, Berlin 10587, Germany (e-mail: s.khalilidehkordi@tu-berlin.de, shuangyang.li@tu-berlin.de, and caire@tu-berlin.de).

P. Fan is with the Institute of Mobile Communications, Southwest Jiaotong University, Chengdu  611756, China (e-mail: pzfan@swjtu.edu.cn).

H. Vincent Poor is with the Department of Electrical and Computer Engineer- ing, Princeton University, Princeton, NJ 08544, USA (e-mail: poor@princeton.edu).
}


}



\maketitle


\begin{abstract}
Next-generation vehicular networks are expected to provide the capability of robust environmental sensing in addition to reliable communications to meet intelligence requirements. A promising solution is the integrated sensing and communication (ISAC) technology, which performs both functionalities using the same spectrum and hardware resources. Most existing works on ISAC consider the Orthogonal Frequency Division Multiplexing (OFDM) waveform. Nevertheless, vehicle motion introduces Doppler shift, which breaks the subcarrier orthogonality and leads to performance degradation. The recently proposed Orthogonal Time Frequency Space (OTFS) modulation, which exploits various advantages of Delay Doppler (DD) channels, has been shown to support reliable communication in high-mobility scenarios. Moreover, the DD waveform can directly interact with radar sensing parameters, which are actually delay and Doppler shifts. This paper investigates the advantages of applying the DD communication waveform to ISAC. Specifically, we first provide a comprehensive overview of implementing DD communications, based on which several advantages of DD-ISAC over OFDM-based ISAC are revealed, including transceiver designs and the ambiguity function. Furthermore, a detailed performance comparison are presented, where the target detection probability and the mean squared error (MSE) performance are also studied.
Finally, some challenges and opportunities of DD-ISAC are also provided.
\end{abstract}


\IEEEpeerreviewmaketitle

\section{Introduction}
\label{Introduction}
\begin{figure}
\centering
\includegraphics[width=0.5\textwidth]{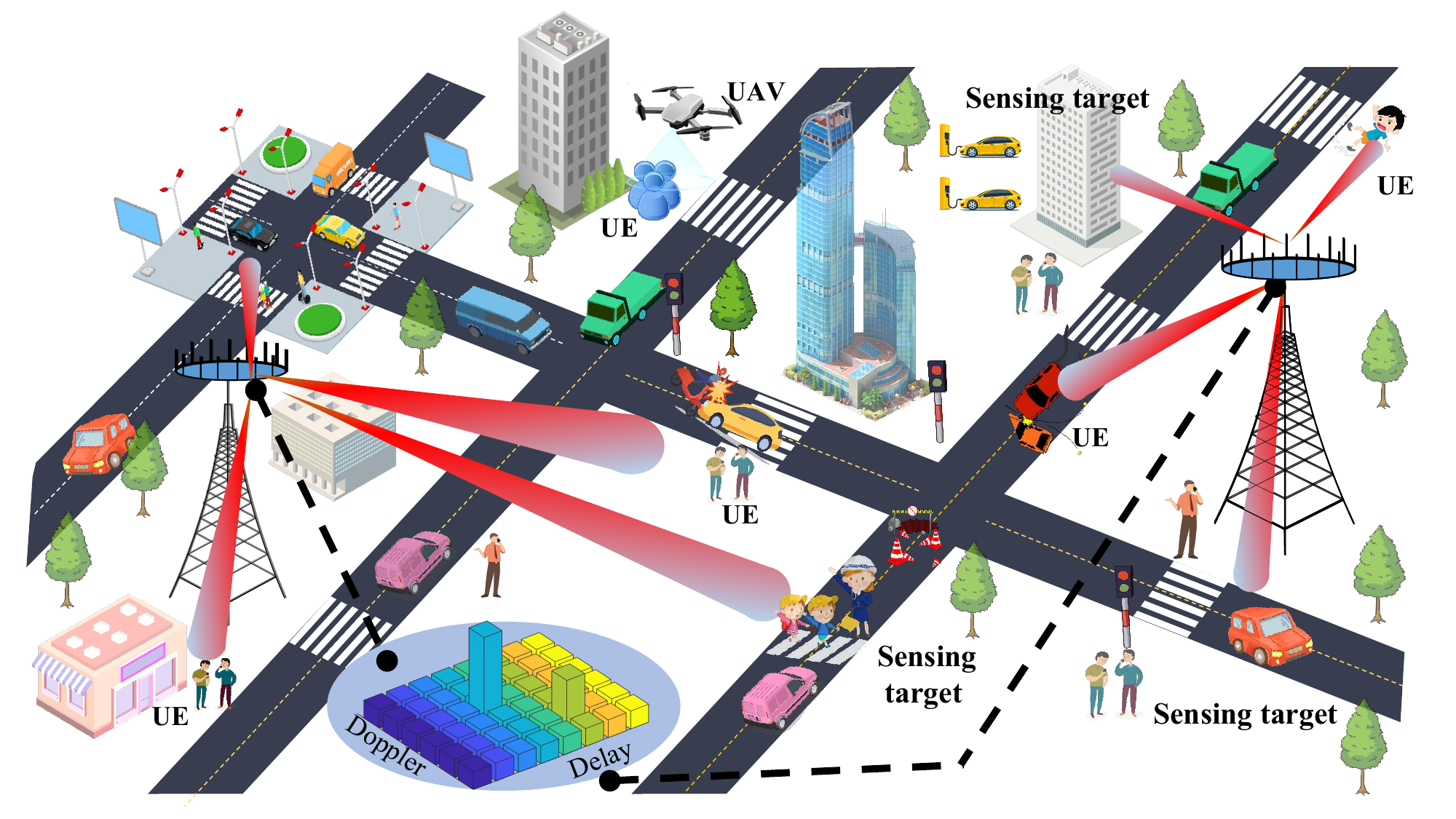}
\caption{DD-ISAC in vehicular networks.}
\label{scenario}
\centering
\end{figure}

Intelligent Vehicular Networks (IVNs) represent an emerging technological trend that are expected to shape modern society in profound ways\cite{paul2016intelligent}. Equipped with an array of advanced sensors and communication devices, vehicles have evolved from traditional transportation modes into intelligent entities. These intelligent vehicles are anticipated to contribute crucially to driving safety, traffic efficiency, and overall commuting experience. The success of IVNs hinges on effectively integrating multidisciplinary technologies, such as environmental awareness, data transmission, and signal processing. Essentially, these coalesce into two core functionalities: communications and sensing, as illustrated in Fig. \ref{scenario}. Particularly, wireless communications technology builds the foundations for user access to the internet and information sharing among vehicles, infrastructures, and pedestrians. On the other hand, the knowledge of surrounding environments, including road conditions, obstacles, and unexpected incidents, is is crucial for enabling functionalities such as autonomous driving. 

Over the past several decades, conventional applications segregated sensing and communication systems, which usually rely on different hardware devices and signal processing frameworks, promoted separately by the radar and communication communities. This duplication of system resources is not sustainable in the long-term development of wireless networks. To this end, a novel trend is emerging that aims to fuse wireless sensing functions into communication devices, in terms of using the same hardware architecture, sharing spectrum resources, and even using unified waveform. This so-called \textit{integrated sensing and communications} (ISAC) technology offers a compelling chance to achieve symbiotic benefits by co-designing sensing and communication functionalities while improving resource efficiency as well as system throughput \cite{liu2022integrated}. 

For the realization of ISAC, an obvious question to ask is: \emph{what kind of waveform should be adopted?} A natural choice is the orthogonal frequency-division multiplexing (OFDM) waveform in the time-frequency (TF) domain, as it has been widely deployed in existing wireless systems. OFDM offers satisfactory performance for communications and sensing relying on the orthogonality between sub-carriers. However, the ultra-reliable information transmission demanded by future IVNs is hard to achieve with OFDM technology since it struggles with high Doppler shifts induced by transceiver movement. To address this issue, orthogonal time frequency space (OTFS), which creatively multiplexes data symbols in the delay Doppler (DD) domain, has been proposed. 
The DD domain multiplexing allows the input-output relation to be fully characterized by the channel response in the DD domain, which is relatively static during signal transmission, compared to its TF domain counterpart \cite{wei2021orthogonal}.
As a result, OTFS is capable of improving communication performance, and is robust to various transmission scenarios, including high-mobility scenarios\footnote{Since the introduction of OTFS in 2017, a variety of OTFS-like waveforms, such as orthogonal delay-Doppler division multiplexing (ODDM) and delay-Doppler alignment modulation (DDAM), have been proposed in the literature. Nevertheless, these schemes hinge on the core concept of multiplexing data symbols in the DD domain. Consequently, they can be categorized under DD communications.}. Furthermore, the inherent connection of the DD parameters to sensing applications enables direct conversion of delay and Doppler shifts into ranges and speeds of sensing targets. Therefore, the DD communication waveform is considered to be a promising candidate for exploiting the full potential of ISAC systems, and has consequently attracted considerable recent attention.

Several studies have investigated DD-ISAC in terms of performance analysis, transceiver design, and comparison with OFDM-ISAC\cite{gaudio2020effectiveness,yuan2021integrated,li2022novel}. However, there is a notable absence of a comprehensive study of DD-ISAC that considers its advantages, performance evaluation, and emerging challenges and areas of research. This article seeks to fill that gap. In particular, we begin with an overview of the basics of DD-ISAC, including an explanation of the mechanics of DD communications and how DD-ISAC is implemented. Following this, we will highlight several benefits related to DD-ISAC, including the unified transceiver design and waveform optimization. Particularly, the ambiguity function of DD-ISAC signals is discussed in detail and compared with that of OFDM. Furthermore, we discuss the target detection and parameter estimation issues that are of core importance in ISAC, and present a comparison between DD-ISAC and OFDM-ISAC in terms of the target detection rate, and estimation mean squared error (MSE).
Finally, we summarize various challenges and opportunities regarding DD-ISAC, and then finish with some conclusions. It is hoped that this work will shed light on the design of DD-ISAC, as well as inspire new ideas related to DD-ISAC.

\section{Preliminaries of DD-ISAC}
The essence of DD-ISAC lies in representing both communication and sensing channels within the same DD domain. The origins of the DD channel model can be traced back to the 1960s, where Bello introduced the wide-sense stationary uncorrelated scattering (WSSUS) channel model~\cite{hlawatsch2011wireless}. The adoption of the WSSUS channel model greatly simplifies the task of communication and sensing, which essentially suggests that the physical attributes, i.e., distance and speed of the channel scatterings remain roughly static during the signal transmission for the given transmitted signal duration~\cite{hlawatsch2011wireless}. Although the theory of the DD domain channel has evolved over the last few decades, most interest has arisen in the radar community since delay and Doppler variables directly represent the range and speed characteristics of moving targets. Its capability for carrying information had been largely overlooked for some time. In fact, representing channel in the DD domain enjoys several advantages, i.e., stability, sparsity, Doppler separability, and compact support, all of which are attractive for communications~\cite{hlawatsch2011wireless}. Having established the historical background of the DD domain channel, we will first explore the fundamentals of DD communications and subsequently discuss the implementation of DD-ISAC.

\subsection{SFFT-based DD Communications}
The main idea of DD communications is to place the information-conveyed data symbols onto a two-dimension (2D) DD plane. Physically, there are not tangible `delay' or `Doppler' resources. These are concepts mathematically derived from time and frequency resources. Given the bandwidth $B$ and time duration $T_f$ of the transmitted signal, we can construct a DD frame composed of $M\times N$ bins, where each bin is occupied by one data symbol. The length and width of a DD bin are defined by the delay resolution $1/B$ and Doppler resolution $1/T_f$, respectively. 
Since the radio waves are transmitted in the time domain, the main task of DD communication transceivers is two-folded: at the transmitter side, to convert data symbols from the DD domain to the time domain signal; and, conversely, at the receiver end, to obtain the signal in the DD domain from the received time domain signal.

It is useful to compare DD communications with conventional TF communications, e.g., OFDM modulation. Considering an OFDM system under critical sampling, where the total TF resources are partitioned into $M\times N$ bins with size $T \times 1/T$. Here $1/T$ is the subcarrier spacing, while $T$ is the time slot duration, where $B=M/T$, and $T_f=NT$, with $M$ being the number of subcarriers and $N$ being the number of time slots, respectively.  Comparing to TF communications, DD communications enable a more dedicated synthesis of TF signals, making it more sensitive to small channel fluctuations in time and frequency, e.g., delay and Doppler.
More specifically, the setup of DD communications indicates that a DD frame can support scatterings with maximum delay of $T$ and Doppler shift of $1/T$, which is sufficient for general underspread channels.

DD communications and TF communications are mathematically connected by the symplectic finite Fourier transform (SFFT). In fact, in the early works on DD communications, the signal transmission was built upon the overlay of OFDM. This implementation is commonly referred to as the SFFT-based implementation.
In general, SFFT-based DD communications relies on a two-stage conversion process where the data symbols in the DD domain are initially transformed into a TF domain signal via the inverse SFFT (ISFFT). This 2D transformation will spread each DD symbol over the entire TF plane. As a result, DD symbols can experience all channel fluctuations in the TF domain, enabling the potential of obtaining the full channel diversity. Then, the standard OFDM modulation is performed, incorporating a TF domain pulse shaping filter, yielding the transmitted signal in the time domain. After passing through the channel, the received signal undergoes multi-carrier demodulation, resulting in the TF domain symbols. The DD domain received symbols are then obtained by applying the SFFT to the TF domain symbols.
Different from TF communications, DD communications enables a channel input-output relation characterized by forms of convolution, e.g., circular convolution, depending on the pulse shapes.
An obvious advantage for SFFT-based DD communications is its compatibility with OFDM modulation, which minimizes the need for modifications of existing hardware equipment. Nevertheless, the introduction of additional processing blocks incurs increased implementation complexity and latency.

\subsection{Zak Ttransform-based DD Communications}
\begin{figure}
\centering
\includegraphics[width=0.45\textwidth]{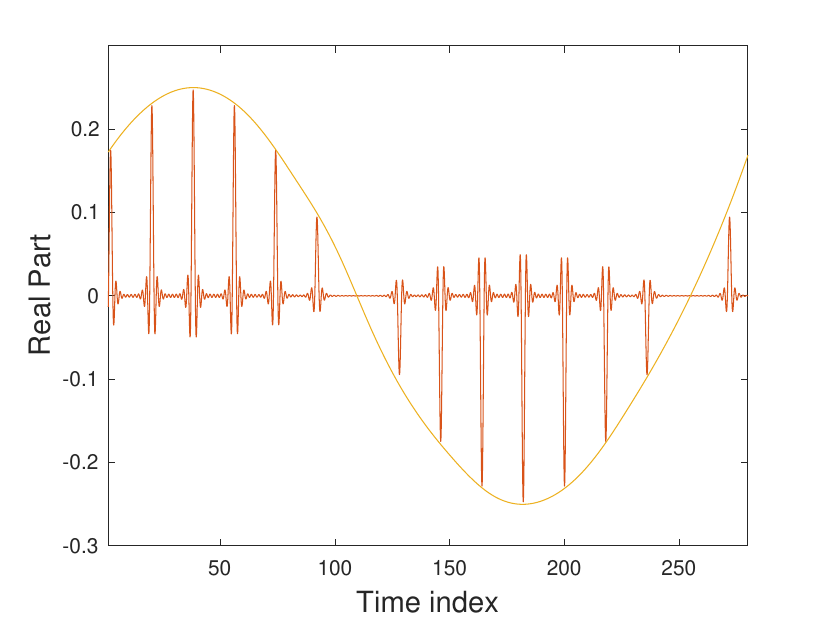}
\caption{A time domain signal as a `pulsone'.}
\label{pulsone}
\centering
\end{figure}

Recent work has shown that the Zak transform can be used to characterize the modulation and demodulation for general DD communications. By leveraging the Zak transform, we are capable of directly converting the time domain signal to the DD domain, bypassing the TF domain processing. The Zak transform, also known as the Gelfand mapping, is an operation that outputs a two-variable function with an input function of one variable, which is essentially a type of Poisson summation. Particularly, the Zak transform has many unique properties, such as quasi-periodicity, which not only simplifies the DD signal representation (The signal in the fundamental rectangle, i.e., a DD region of $\tau  \in \left[ {0,T} \right)$ and $\nu  \in \left[ {0,1/T} \right)$, fully characterizes the signal in the entire DD domain.), but also gives rise to special signal structures in both time and frequency. To place the DD symbols on a carrier, the DD domain basis functions (carriers) can be constructed in a way that aligns with the properties of the Zak transform. Specifically, it was shown in~\cite{Shuangyang2023Globecom} that the DD domain basis function is quasi-periodic globally, while twisted-shifted locally within the fundamental rectangle. {Such a basis function turns out to be the shape of a ``pulsone" in both time and frequency as shown in Fig. \ref{pulsone}}. In this figure, the pulsone in time is a train of time-delayed pulses modulated by a signal tone determined by the Doppler. Correspondingly, it can be shown that the pulsone in frequency is a train of frequency-shifted pulses modulated by a signal tone determined by the delay. This symmetrical treatment in time and frequency allows a full exploitation of the channel dynamics thereby improving the communication performance.

In line with the properties of the Zak transform, DD modulation can be implemented digitally using the inverse discrete Zak transform (IDZT), followed by filters and windows for converting it into time-continuous waveform with approximately limited TF resources. Similarly, at the receiver side, we can apply matched filters and windows to discretize the signal, followed by a DZT, to yield the received DD symbols.
We shall refer to this type of implementation as the Zak transform-based implementation. In fact, both SFFT-based and Zak-based DD communications will produce similar input-output relations depending on the underlying filters and windows. 
Nevertheless, the Zak-based DD communication requires only an $N$-point Doppler-time finite Fourier transform with a symbol-wise interleaver, leading to a lower implementation complexity compared to the SFFT-based method. Furthermore, Zak-based DD communications enjoys many practical benefits. For example, the information carrier can be directly designed in the DD domain without leveraging TF domain pulse shaping. Therefore, Zak-based DD communication does not suffer from the pulse discontinuity issue, which in return improves the out-of-band emission. 
However, the filter and window design for 
Zak transform-based implementation are of critical importance for achieving orthogonality and preserving the DD communication features. This important topic will be discussed in Section-III of this paper.




\subsection{Implementation of DD-ISAC}
\begin{figure}
\centering
\includegraphics[width=0.5\textwidth]{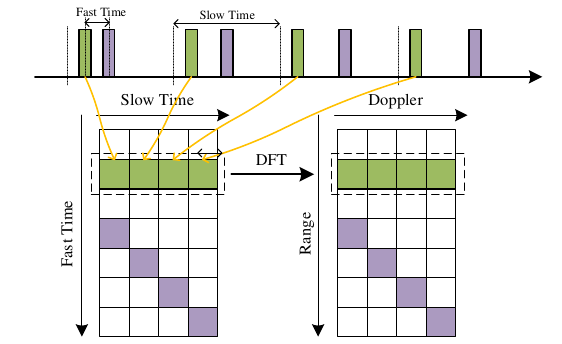}
\caption{Illustration of the connection between range-Doppler matrix calculation and DD symbol demodulation.}
\label{DZT}
\centering
\end{figure}

Since the establishment of DD communications, its capability for radar sensing has attracted significant interest. Intuitively, DD parameters, which can describe a target's state, align perfectly with the sensing tasks. To shed light on this, we can consider radar sensing.  Typical radar sensing studies revolve around the calculation of the range-Doppler matrix. This computation involves applying the discrete Fourier transform (DFT) to the slow-time axis of the fast-time slow-time matrix. Here, fast time corresponds to the sampling rate, which relies on the available bandwidth of the instantaneous pulse, while slow time represents the sampling interval between successive pulses. After matched-filtering the time domain sensing signal and sampling at an appropriate rate, the samples can be reorganized into a 2D matrix, as shown in Fig. \ref{DZT}. Then, performing a DFT along the slow-time axis produces the range-Doppler matrix, which 
essentially indexes the matrix elements by delay and Doppler (the range and delay as a one-to-one correspondence related to the signal propagation speed).
Recall the earlier discussions on the demodulation of Zak-based DD communications. The time domain sequence after matched-filtering is converted to a matrix in the DD domain via the DZT. In fact, the DZT can be regarded as the combination of a symbol-wise interleaver and a DFT applied along the time axis. Therefore, it is straightforward to conclude that the derivation of range-Doppler matrix is exactly identical to the demodulation process in DD communications \cite{Kecheng2023ICC}. The only difference lies in that radar sensing aims to extract the state parameters from the DD matrix, whereas communications seeks to detect symbols conveying information.

Next, we will show how sensing and communication sub-systems can collaborate with each other to achieve coordination gain. As discussed in \cite{yuan2021integrated}, the DD and angular parameters extracted from the sensing signals can facilitate inference about the communication channel, which is referred to as `sensing-assisted communications'. For instance, suppose a base station (BS) transmits a dual-functional DD-ISAC signal. Assuming a mono-static radar setup, the sensing echo is received at the BS. Through advanced estimation techniques, delay and Doppler shifts associated with moving targets can be obtained, which are adopted for constructing the topology of the mobile network. Subsequently, the network topology in the next few time instants can be predicted, allowing the prediction of the DD communication channels. In scenarios with line-of-sight (LoS) dominated channel model, both the path loss and phase offset of the channel can be accurately predicted and compensated at the BS, leading to a very low cost at the user side. When dealing with the generalized multi-path channel case, although perfect prediction of the communication channel remains elusive, the knowledge of DD shifts can provide a known interference pattern in the DD domain, which can reduce the channel estimation overhead. 

Moreover, the DD domain channel intrinsically reveals the underlying wireless propagation environment. Hence, the estimates of the communication channel can serve a dual purpose: not only facilitating communications but also providing the capability of coarse environmental sensing, termed `communications-assisted sensing'\cite{yuan2022orthogonal}. Following standard communication processing, channel estimation precedes data detection. The estimated delay and Doppler indices directly mirror the physical attributes linked to in-environment scatterings. To illustrate, consider an environment with a BS and a moving vehicle. At each time instant, the vehicle estimates the downlink DD channel and obtains the delay and Doppler shifts associated with different paths. Through straightforward manipulations, the distance of the BS-scattering-vehicle path can be measured based on the estimated delay. Having obtained the range measurements over several consecutive instants, a standard elliptical localization problem can be formulated where the BS and vehicle are the foci of the ellipses. In addition, the Doppler shift containing angular information can help differentiate measurements with respect to different paths. 


\section{Benefits of DD-ISAC}
Implementing ISAC in the DD domain is a natural extension of the current TF domain ISAC schemes, e.g., OFDM-based ISAC, and it has advantages in many aspects. As the 
benefits of DD communications have been widely discussed in the literature~\cite{wei2021orthogonal}, in this section, we will focus on the advantages of sensing in the DD domain and DD domain integration of the two functionalities.

Perhaps, the most obvious advantage of DD-ISAC is the unified transceiver design. Conventional ISAC designs may require different modules to achieve communication and sensing functionalities at the same time. However, this may be difficult for the hardware design, because of limited chip size and increased complexity. Against this background, DD-ISAC may be a better solution for ISAC because of the overlay between the range-Doppler matrix calculation and demodulation of DD signals, e.g., the IDZT. 
This may be important in fully-integrated ISAC systems, where the same waveform is transmitted for both communication and sensing functionalities.
In this case, a unified transceiver can be applied for both functionalities, which reduces the hardware expense. Particularly, to achieve good sensing performance, the OTFS-based receiver can first perform OTFS demodulation, followed by a matched-filter implemented digitally, which is known as `pulse compression'. A scheme of this kind appears in~\cite{Kecheng2023ICC}, which shows promising sensing performance under various channel conditions.   

Beyond the implementation convenience, DD-ISAC may enjoy improved sensing performance by exploiting the unique features of its ambiguity function. According to~\cite{Shuangyang2023Globecom}, the ambiguity function of the DD basis function is of the following form   
\begin{align}
      &\int_0^T \int_0^{\frac{1}{T}} \Phi _{{\rm{DD}}}^{\tau_2,\nu_2}\left( \tau ,\nu  \right){{\left[ {\Phi _{{\rm{DD}}}^{\tau_1,\nu_1}\left( {\tau ,\nu } \right)} \right]}^*}{\rm{d}}\nu {\rm{d}}\tau  \notag\\
    =& {e^{j2\pi {\nu _2}\left( {{\tau _1} - {\tau _2}} \right)}}{A_\Phi }\left( {{\tau _1} - {\tau _2},{\nu _1} - {\nu _2}} \right) \label{DD_MF_basis_functions}.
\end{align}
In~\eqref{DD_MF_basis_functions}, $\Phi _{{\rm{DD}}}^{\tau_0,\nu_0}\left( \tau ,\nu  \right)$ is the DD basis function associated with the delay index $\tau_0$ and Doppler index $\nu_0$, which carries the DD domain symbol, while ${A_\Phi }\left( {\Delta \tau,\Delta \nu} \right)$ is the ambiguity function of the DD domain prototype pulse $\Phi _{{\rm{DD}}}\left( \tau ,\nu  \right)$ with respect to delay shift $\Delta \tau$ and Doppler shift $\Delta \nu$, where $\Phi _{{\rm{DD}}}^{\tau_0,\nu_0}\left( \tau ,\nu  \right)$ is connected to $\Phi _{{\rm{DD}}}\left( \tau ,\nu  \right)$ via linear shifting and phase-rotation. Essentially,~\eqref{DD_MF_basis_functions} states that the connection between the ambiguity function and the DD basis function is simply a two-dimensional correlation. Note that the DD basis function can be fully localized in the fundamental rectangle, which is a region of DD components defined by $\tau \in [0,T)$ and $\nu \in [0,1/T)$, without violating Heisenberg's uncertainty principle. This is because the DD basis function is only localized in the fundamental rectangle but quasi-periodically repeated across the entire DD domain, which can be easily verified by taking the Zak transform of periodic signals. With such a basis function, its ambiguity function is also fully localized locally but is periodic globally.
Therefore, by carefully selecting $T$ according to the channel conditions, radar sensing using DD signals can enjoy an improved signal-to-noise ratio (SNR) and avoid delay and Doppler ambiguities.

\begin{figure}
\centering
 \includegraphics[width=0.48\textwidth]{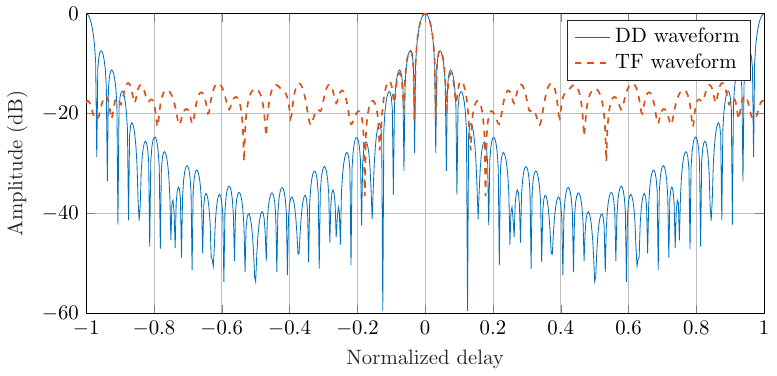}
\caption{Zero-Doppler cut of the ambiguity function.}
\label{zero_Doppler}
\centering
\end{figure}

\begin{figure}
\centering
\includegraphics[width=0.48\textwidth]{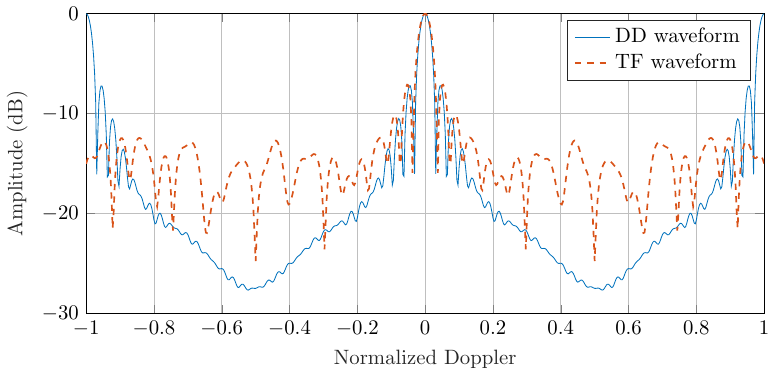}
\caption{Zero-delay cut of the ambiguity function.}
\label{zero_delay}
\centering
\end{figure}

Unfortunately, the aforementioned DD basis function only exists given infinite time and frequency resources, which cannot be achieved in practice, and the accuracy of localization decreases with limited time and frequency resources. Consequently, careful pulse shape design is important in practice. It was shown in~\cite{Shuangyang2023Globecom} that truncating localized DD basis functions with periodic or orthogonal windows in frequency and time introduces approximate DD orthogonalities with respect to the delay and Doppler resolutions~\cite{Shuangyang2023Globecom}, where the ratio between the main lobe and the side lobe of the ambiguity function depends on the underlying shape of windows. This observation provides clear insights into the DD pulse shape problem, i.e., the pulse shape design
reduces to the window design with limited bandwidth and time resources for highly structured signals. More specifically, the ambiguity function of windowed DD basis functions can be decomposed into the summations of ambiguity functions of the underlying window functions. These properties greatly simplify the DD pulse shaping problem by allowing specific objective formulations according to the ambiguity function requirements. 
Although at the current stage, the DD pulse shaping problem for ISAC has not been fully solved, preliminary results in~\cite{Shuangyang2023Globecom} have demonstrated significant potential of DD-ISAC. In Fig.~\ref{zero_Doppler} and Fig.~\ref{zero_delay}, we present the zero-delay and zero-Doppler cuts of the ambiguity function for pulse-shaped DD and TF waveforms, i.e., OFDM, where the underlying pulse is the root raised cosine (RRC) pulse with a roll-off factor $\beta=0.3$ and we have $T=1$. Specifically, we consider  $M=N=32$, and both waveforms are implemented by using a discrete multitone (DMT) structure. Particularly, all information symbols of the TF waveform are randomly chosen from a QPSK constellation, while for the DD waveform, we assume that all the information symbols are zeros except for the symbol at the first delay/Doppler bin, where the total transmitted power for the two waveforms are normalized to be the same. Note that, with a single non-zero information symbol for the DD waveform, the transmitted DD signal after modulation is essentially the same as the DD domain basis function discussed above, which can be achieved by inserting a sufficiently large guard space around the DD information symbol in practice.
Based on  Fig.~\ref{zero_Doppler} and Fig.~\ref{zero_delay}, we observe that the DD waveform has an ambiguity function that decays rapidly, while maintaining a roughly periodic structure with respect to $T$ and $1/T$. In comparison, the ambiguity function of the TF waveform fluctuates at different delay and Doppler values. This is because the ambiguity function of the TF waveform relies on the independence among different transmitted symbols. From the two figures, we can conclude that the DD waveform is better suited for sensing tasks in comparison to the TF waveform. However, the value of $T$ and $1/T$ need to be carefully chosen in order to avoid delay and Doppler ambiguities.

\section{Performance Evaluation of DD-ISAC}
In this section, we provide an overview of the target detection and estimation aspects that comprise the \textit{sensing} functionality of DD-ISAC system. Particularly, we will compare these aspects with those of the TF waveform, in order to provide a comprehensive understanding.

\subsection{Target Detection}\label{disc_param}
\begin{figure}
\centering
\includegraphics[width=0.4\textwidth]{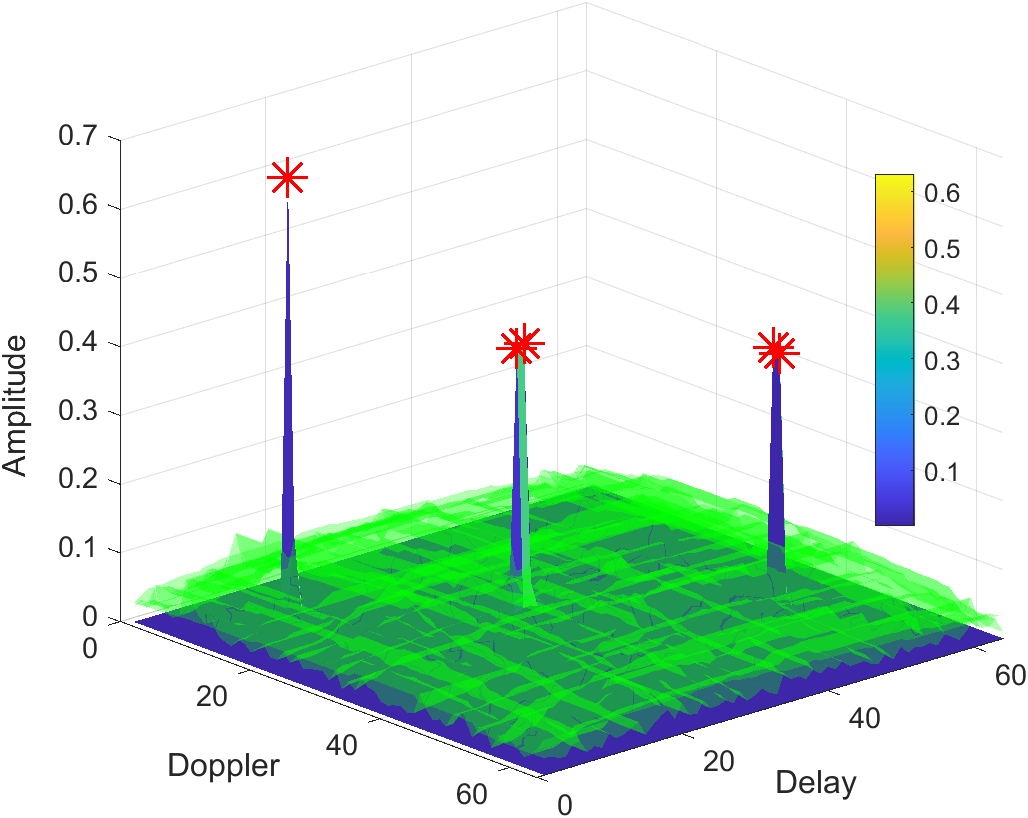}
\caption{Detection with an adaptive threshold (shown as the transparent layer).}
\label{dd_detec}
\centering
\end{figure}

Consider an unknown number $P$  of targets in the sensing Field of View (FoV) of the ISAC system. For simplicity, assume that these targets can be modeled as single scatterers. Since the number of targets is initially unknown, a \textit{detection} step is required to distinguish between targets, clutter, and noise. To achieve this, we formulate the target detection as a standard Neyman-Pearson hypothesis testing problem. Under this framework, at each step we have hypotheses $\Hc_0$ and $\Hc_1$ corresponding to the absence or presence of a $p$-th target only. This means that when detecting a target and estimating the relevant parameters, the targets already detected are assumed to have already been canceled from the received signal, whereas the contributions of the remaining targets are modeled as additional noise. 
The observation under the two hypotheses is given by 
\begin{align} \label{eq:Hypo_test}
    \yv = \begin{cases}
    \wv & \;\; \text{under $\Hc_0$}\\
    \Am(\mathring{\Theta}_p) \xv  + \wv
    & \;\; \text{under $\Hc_1$}\,.
    \end{cases}
\end{align}
where $\mathring{\Theta}_p$ denotes the \textit{true} parameters of the $p$-th target, $\Am$ is the radar channel matrix, containing the channel gain, delay-Doppler, and possibly beamforming effects, and $\wv$ is the noise term. The Neyman-Pearson hypothesis testing problem above, for which the solution that maximizes the detection probability subject to a bound on the false-alarm probability is given by the (Generalized-) Likelihood Ratio Test (LRT)~\cite{VPoor}. Here a threshold determines the tradeoff between detection and false-alarm probabilities for target detection. Since the statistics of noise and interference in realistic scenarios are generally not known, an adaptive threshold function according to the Constant False Alarm Rate Detection (CFAR) approach can be used. In particular, the Ordered Statistic (OS) -CFAR method, is known to provide good performance in those scenarios \cite{Dehk_TWC}. 
As an example, Fig.~\ref{dd_detec} shows the performance of the OS-CFAR technique applied to the DD radar image in a multi-target scenario. 
\begin{figure}[h]
\centering
\includegraphics[width=0.48\textwidth]{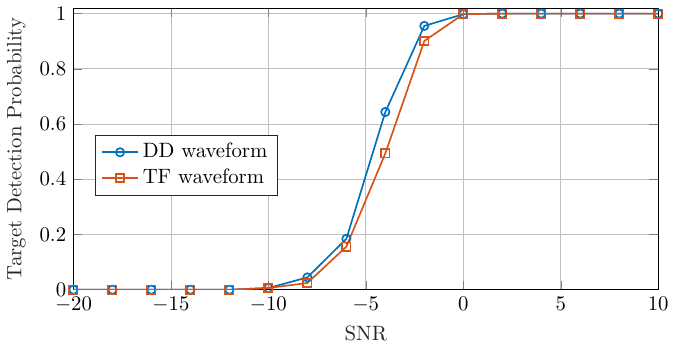}
\caption{Probability of detection comparison of the two waveforms for a single target using the adaptive OS-CFAR thresholding technique, with identical parameters for both waveforms.}
\label{fig:PD_vs_SNR}
\end{figure}

\subsection{Target Parameter Estimation}\label{estm_param}
Parameter estimation can be carried out as a side-product of the LRT-based detection above. More precisely, by defining a discrete grid over the parameter search space (e.g. delay, Doppler, angle)  and evaluating the likelihood ratio at each grid point, the grid cells passing the defined threshold indicate the target parameters. In multi-target cases, a target located closely may create `masking' effects. This is especially significant when a target is a much stronger reflector than adjacent targets. Hence, a sequential detection and estimation framework of targets was proposed in \cite{Dehk_TWC}, where after each target is detected and its parameters are estimated, a Successive Interference Cancellation (SIC) routine is carried out. This step effectively cancels out the contribution of a detected target from the received signal and proceeds to the detection of the next target.

\begin{figure}[h]
\centering
\includegraphics[width=0.48\textwidth]{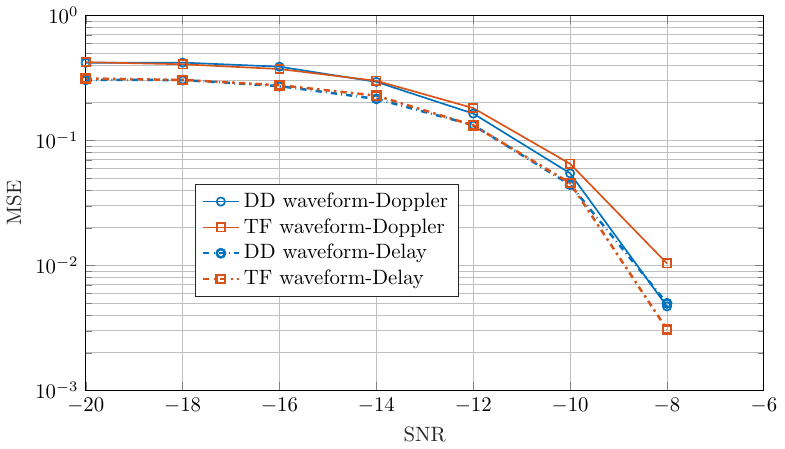}
\caption{MSE comparison of DD and TF waveforms for parameter (delay and Doppler) estimation of a single target.}
\label{fig:Delay_Dopp_MSE}
\end{figure}


\subsection{Performance Comparison with ISAC in the TF domain}
In this subsection, we will present numerical results discussing the sensing benefits of DD-ISAC in comparison to OFDM-based ISAC. We argue that DD-ISAC has the potential to outperform OFDM-based ISAC thanks to its advantage in terms of the ambiguity function. Particularly, the ambiguity function of DD-ISAC shows a robust and `spike-like' response for normalized delay and Doppler. This observation clearly indicates good sensing performance when applying target detection and parameter estimation based on a matched-filtering type of receiver. Without loss of generality, both DD and TF waveforms are pulse shaped by the RRC pulse with roll-off factor $\beta=0.3$ via the DMT structure.

We present the numerical results of the target detection probability based on matched-filtering in Fig.~\ref{fig:PD_vs_SNR}, where we consider $M=N=16$ and the number of targets is set to be $1$. As can be observed from the figure, DD-ISAC shows a better detection probability than the OFDM-based ISAC using the CFAR approach. This can be attributed to the `less noisy' radar image in the DD domain produced by the DD waveform, as observed from Fig. \ref{zero_Doppler} and Fig.  \ref{zero_delay}. Specifically, both DD and TF waveforms start to work when the SNR is higher than $-10$ dB, while the detection probability achieves almost one with SNRs higher than $0$ dB. 

In Fig.~\ref{fig:Delay_Dopp_MSE}, the MSE performance of parameter estimation is shown, where $M=N=16$ and a single target is considered. Specifically, we calculate the associated MSE based on the detected target. Furthermore, if the CFAR fails to detect the target, we simply take the corresponding delay and Doppler with the highest response after matched-filtering as the estimator output. From the figure, we notice that both DD and TF waveforms provide similar MSE performance in general. This is because the considered frame length is sufficiently long to ensure a satisfactory auto-correlation performance of the transmitted symbols, such that the parameter estimation based on matched-filtering is good enough. 

We highlight that we only adopted a single target ISAC in the above examples because it is the simplest and most common scenario for sensing. However, in practice, the wireless channel for sensing can be quite complex with many unintended scatterers. In this case, the DD waveform may provide further performance improvement for sensing in comparison to the TF waveforms, thanks to its `spike-like' ambiguity function.

\section{Challenges and Opportunities}

\subsection{Waveform Design in the Delay Doppler Domain}
Some preliminary results show that a single symbol placed in the DD domain might be sensing-optimal from the AF perspective. Nevertheless, rigorous mathematical validation is needed. Starting from this point, a very interesting topic is DD-ISAC waveform design. Since communication systems lean towards randomized signals while sensing favors deterministic ones, there is an inherent deterministic-random trade-off for ISAC waveform design. When our aim is to improve sensing performance, a possible solution is to optimize the transmitted DD sequence that closely resembles the optimal sensing sequence in the expectation sense. In contrast, if we take the communication performance as the primary goal, a water-filling type of DD precoding design should be adopted to achieve channel capacity, which however inevitably affects the correlation of the DD sequences and degrade the sensing performance. Hence, `optimal' DD-ISAC waveforms vary based on specific applications, drawing considerable attention nowadays. 

\subsection{Performance Analysis Framework and Algorithmic Design}
Existing works study the performance of ISAC using channel coding for communications and using either rate-distortion theory or the Cram\'er-Rao bound for sensing. The communication task is very well understood while the sensing task, especially the performance trade-off between communication and sensing is less understood. Based on R\'enyi's pioneering work~\cite{renyi1961problem}, the ideas from 20 questions estimation~\cite{zhou2021resolution} might be adopted, which converts the sensing and communication tasks to continuous and discrete random variable estimation problems. Specifically, the sensing task of DD-ISAC, which focuses on the estimation delay, Doppler shifts, and reflection coefficients, can be converted to a 20 questions estimation problem of multiple targets in multi-dimensions, as studied in~\cite{zhou2022multiple}. As a result, the theoretical results can be adapted to yield bounds on the estimation error for targets. Using the connection of the sensing task of ISAC to 20 questions estimation, the performance bounds on the estimation error could be obtained, which depend on the sensing channel via the mutual information term and an additional dispersion term that characterizes the variance of the channel. Correspondingly, the transmission performance, defined as the maximal transmission rate subject to constraints on transmit power and error probability is a function of the communication channel via similar mutual information and dispersion terms. In case of quasi-static fading, both performance criteria for sensing and communication essentially reduce to mutual information over the corresponding channels.

Although we can potentially use 20 questions estimation to obtain theoretical bounds on the sensing performance, another concern is the complexity of algorithms. Some preliminary results rely on random coding with maximum likelihood decoding, which has exponential complexity with the signal length. However, no practical systems could support such high complexity due to power and latency constraints. Thus, another challenge is how to design low complexity algorithms for time and resource efficient target sensing.

To solve this problem, we can resort to the construction of low-complexity query procedures for 20 questions estimation. Specifically, since any channel code can be used to construct a query procedure for 20 questions estimation, we can use practical low-complexity codes such as LDPC or polar codes to construct query procedures and adapt the query procedure to design the signal and the sensing decoder.

\subsection{Integration with RIS and Near-Field Communications}
The reconfigurable intelligent surface (RIS) technique aims at modifying the phase of the transmitted signal to enhance the communication performance and save on hardware costs and energy consumption. Relying on the DD-ISAC framework, one cannot only improve the communication performance but also can attain a better sensing performance as well. In particular, the adoption of of RIS technology can directly modify the DD response of the physical channel, which in return improves the sensing performance.

As a further comment, near-field communication has emerged as a topic of interest in the community recently. In contrast to traditional far-field communications, the range distance information will be exploited for beamforming and multiple access design in near-field communication, which actually reflects the delay parameter. Therefore, incorporating DD signaling into near-field communications is interesting and deserves further investigation.

\section{Conclusions}
\label{CONCLUSION}
In this article, we have explored the innovative idea of integrating sensing and communications in a unified DD domain. The consideration of the DD domain, in contrast to the traditional TF domain domain brings new opportunities for advancing ISAC. We began with an in-depth discussion of the foundations of DD communications and extended it to the implementation DD-ISAC. 
Then, the benefits of DD-ISAC in terms of implementation and ambiguity functions were discussed. Furthermore, a detailed study of target detection and parameter estimation was presented, where the comparison with OFDM-based ISAC was also presented.
Finally, various promising research directions were highlighted. We hope this tutorial serves as a useful forward-looking guide to research in DD-ISAC.

\bibliographystyle{IEEEtran}
\bibliography{ref}

\end{document}